\documentclass[a4paper,12pt]{article}
\usepackage{bbm}
\usepackage[utf8]{inputenc}
\usepackage{amssymb,amsmath,amsfonts}
\usepackage{mathrsfs}
\usepackage[colorlinks,linkcolor=blue,citecolor=Green,urlcolor=blue]{hyperref}
\usepackage{color}
\usepackage{graphicx}
\usepackage{hyphenat}
\usepackage[dvipsnames]{xcolor}
\usepackage{wrapfig}
\usepackage{empheq}
\usepackage{textcomp}
\usepackage[caption=false]{subfig}
\usepackage{rotating}
\usepackage{wrapfig}
\usepackage{pdfpages}
\usepackage{verbatim}
\usepackage{setspace}
\usepackage{array}
\usepackage{caption}
\usepackage[pdf]{pstricks}
\usepackage{upgreek}
\usepackage{cmll}
\usepackage{latexsym}
\usepackage{braket}
\usepackage{cite}
\usepackage{epsfig}
\usepackage[left=2.5cm,top=3cm,right=2.5cm,bottom=3cm,bindingoffset=0cm]{geometry}
\usepackage[titletoc,page]{appendix}
\usepackage{multicol}
\usepackage{youngtab}

\numberwithin{equation}{section}

\begin{document}

\title{\bf  
Probabilistic Parallels in the Classical Limit of
Quantum Mechanical Models
}  
\author{
Raghunathan Ramakrishnan$^{(1)}$ \\[13pt]
{\em \footnotesize
$^{(1)}$Tata Institute of Fundamental Research, Hyderabad 500046, India
}\\
\\[-5pt]
\\[1pt] 
{  \footnotesize  \texttt{ramakrishnan@tifrh.res.in} }  
}
\date{}

\maketitle

\begin{abstract}
At large quantum numbers, the probability densities for 
particle-in-a-box or simple harmonic oscillator 
converge to the classical result upon coarse-graining the 
quantum mechanical probability densities by introducing a finite 
resolution in the measurement of the particle's position. This resolution in the position can be related to the resolution of the secondary total angular momentum quantum number ($m$) when interpreting the probabilistic outcomes of the Stern--Gerlach-type thought experiments for large values of the angular momentum quantum numbers ($j$).
\end{abstract}

%\newpage  
%\tableofcontents
%\newpage

\section{Introduction}  
Understanding the classical limit in quantum mechanics sheds light on how to interpret the wave-like behavior of quantum probability distributions at large quantum numbers. For systems such as the particle-in-a-box or simple harmonic oscillator, an elegant approach presented in~\cite{Liboff,LiboffQM} uses local averaging:
\begin{eqnarray}
    \langle P_{\rm QM}\rangle(x) & = & \frac{1}{2\epsilon} 
    \int_{x-\epsilon}^{x+\epsilon} dx^\prime  | \psi_n(x^\prime) |^2,
    \label{eq:local_average}
\end{eqnarray}
where $\epsilon$ represents an arbitrarily small interval around $x^\prime=x$. 
In the case of angular momentum, the relevant probabilities correspond to the square of the reduced Wigner functions that are also known as the elements of the d-matrix:
\begin{eqnarray}
    P_{m^\prime m}^{(j)} \left( \beta \right) = \left| d_{m^\prime m}^{(j)} \left( \beta \right) \right|^2,
    \label{eq:prob_wigner}
\end{eqnarray}
where 
\begin{eqnarray}
    d_{m^\prime m}^{(j)} ( \beta) & = & 
    \langle j, m^\prime | \exp\left( \frac{-iJ_y\beta}{\hbar} \right) | j, m \rangle 
    =  \langle j, m^\prime | j, m; \beta \rangle. 
\end{eqnarray}
The Wigner function, depending on the angle $\beta$ and quantum numbers $j$, $m$, and $m^\prime$, is given as follows~\cite{SakuraiQM}
\begin{eqnarray}
    d_{m^\prime m}^{(j)} \left( \beta \right) & = &
    \sum_{k=\max(0,m-m^\prime)}^{\min(j+m,j-m^\prime)} (-1)^{k-m+m^\prime} 
    \frac
    {
    \sqrt{ \left(j + m \right)! \left(j - m \right)! \left(j + m^\prime \right)! \left(j - m^\prime \right)! }
    }
    {\left( j + m - k \right)! k! \left( j - k - m^\prime  \right)! \left( k - m + m^\prime  \right)! }  \nonumber \\
    & & \times
    \left( \cos \frac{\beta}{2}  \right)^{2j-2k+m-m^\prime}
    \left( \sin \frac{\beta}{2}  \right)^{2k-m+m^\prime}.
    \label{eq:WignerFunction}
\end{eqnarray}
The index and limits of the 
summation in Eq.~\ref{eq:WignerFunction} ensures that 
the arguments for the factorials in the denominators are $\ge0$.
Another version of this formula is~\cite{LandauQM} 
\begin{eqnarray}
d_{m^\prime m}^{(j)} \left( \beta \right) & = &
\sqrt{\frac{(j+m^\prime)!(j-m^\prime)!}{(j+m)!(j-m)!} }
\left( \cos \frac{\beta}{2}\right)^{m^\prime+m}
\left( \sin \frac{\beta}{2}\right)^{m^\prime-m} 
\Phi_{j-m^\prime}^{m^\prime-m,m^\prime+m} \left( \cos \beta \right),
\label{eq:WignerFunction_Landau}
\end{eqnarray}
where $\Phi_n^{a,b}$ are Jacobi polynomials. 

To interpret Eq.~(\ref{eq:prob_wigner}), let us define the state ket,
$ | j, m  \rangle $, to denote the state of
a quantum mechanical angular momentum vector $\vec{J}$ of length $\sqrt{ j(j+1) }$ with the component along $z$-axis known to be $m$, in units of $\hbar$. The state ket
$ | j, m; \beta  \rangle $ denotes the state of $\vec{J}$ with its length measured along the 
$z^\prime$-axis, which is the $z$-axis rotated through an angle $\beta$ about the $y$-axis. Wigner function gives the probability amplitude for a rotated state ket, $ | j, m; \beta  \rangle $, along the unrotated basis kets, $| j, m^\prime \rangle $. 

The purpose of this pedagogical note is to relate the ``small interval'' ($\epsilon$) in Eq.~(\ref{eq:local_average}) to uncertainty in the measurement of position, and 
present the classical limit of selected cases of 
    $P_{m^\prime m}^{(j)} \left( \beta \right)$ (denoted as Wigner probabilities)
    on the same footing as  
    probabilities encountered in the simpler
    model problems particle-in-a-box or simple harmonic oscillator.

\section{Wigner Probabilities as Stern--Gerlach Outcomes}
\begin{figure}[!htpb]
\begin{center}
\includegraphics[width=1.0\linewidth]{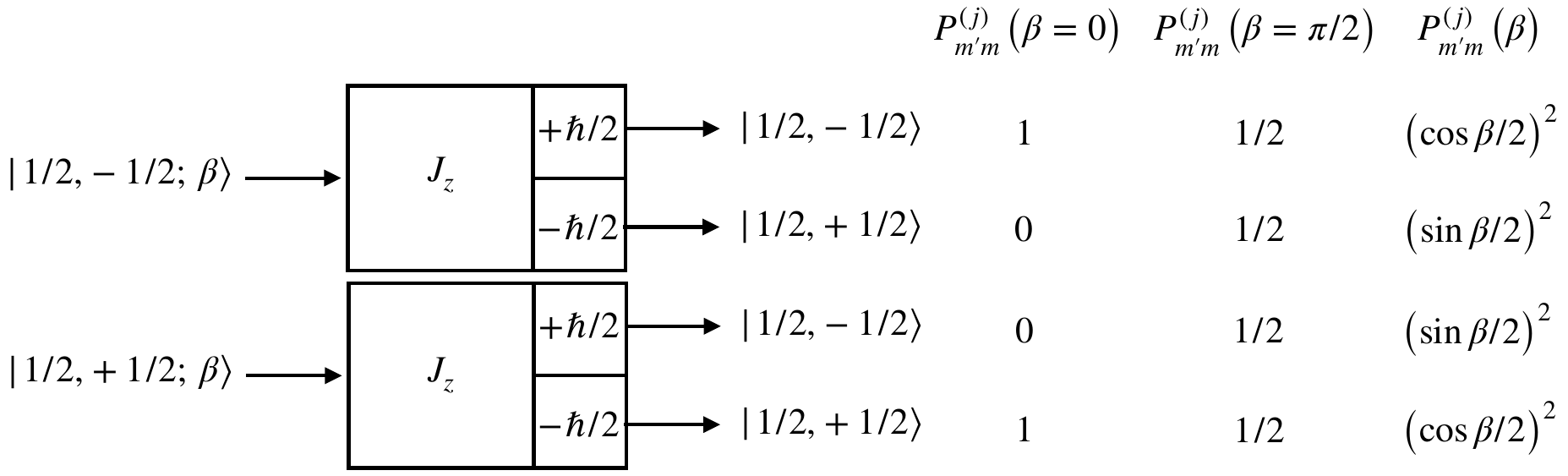}
\caption{
\footnotesize {Probabilistic outcomes of SGE for $j=1/2$.}
}
\label{fig:SGE_half}
\end{center}
\end{figure}

\begin{figure}[!hbpt]
\begin{center}
\includegraphics[width=1.0\linewidth]{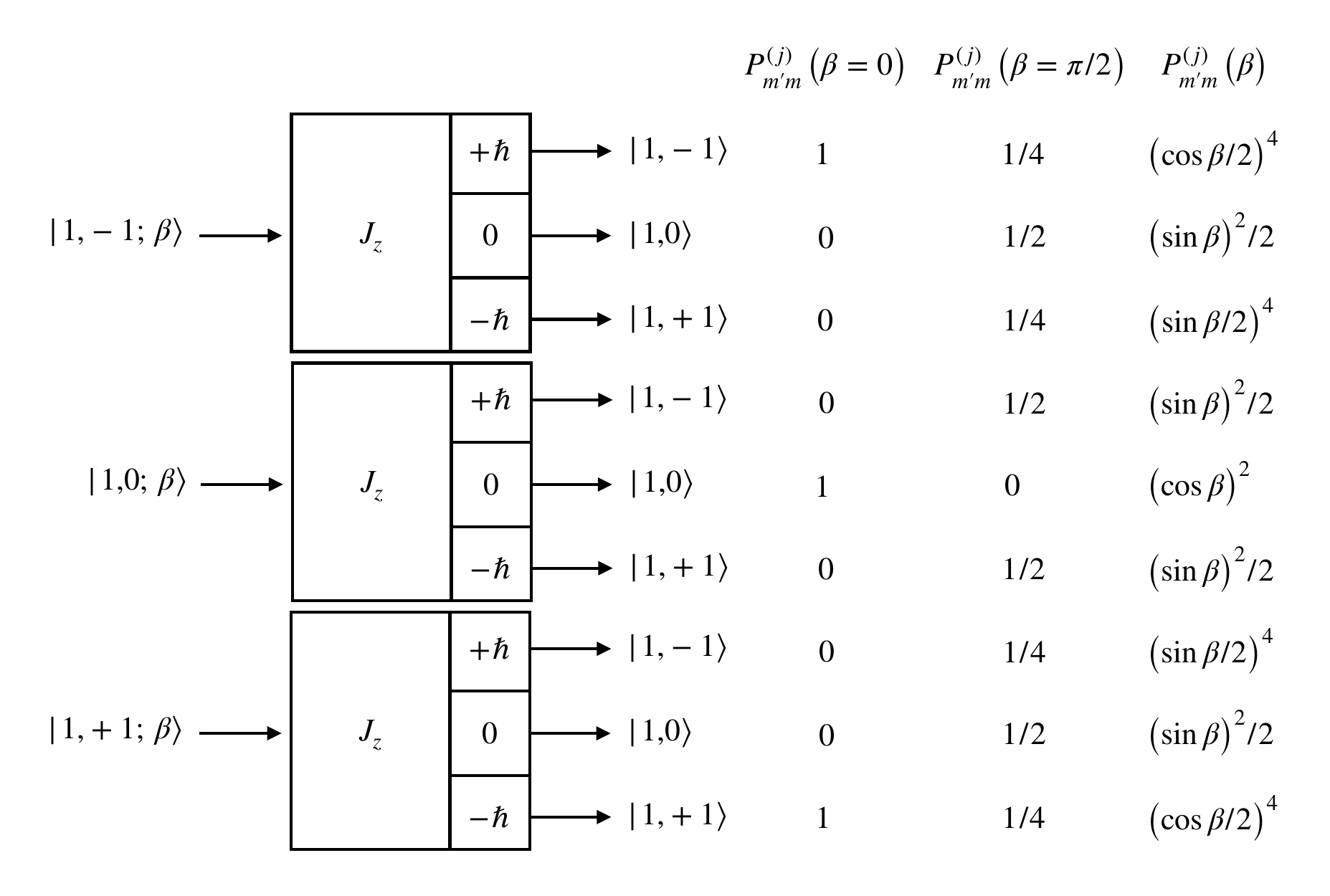}
\caption{
\footnotesize {Probabilistic outcomes of SGE for $j=1$.}
}
\label{fig:spin_one}
\end{center}
\end{figure}

In the modern pedagogy of quantum mechanics~\cite{SakuraiQM}, 
the postulates of quantum mechanics and
the matrix representations of the angular momentum operators are  
presented axiomatically to explain the probabilistic outcomes of the 
classic Stern--Gerlach experiment (SGE) for measuring the spin-projection ($s_z$)
and its hypothetical variants applied to the total angular momentum, $J_z$.  
When $\beta=0$, the state kets, $|j,m\rangle$, correspond to the basis kets
$| 1/2, \pm 1/2\rangle=|\pm z\rangle$ that are the eigenkets of $s_z$ . For
$\beta=\pi/2$ the state kets correspond to the eigenkets of $s_x$,
$| 1/2, \pm 1/2; \pi/2\rangle=|\pm x\rangle$. 
In the case of $j=1/2$, the equal probabilities of $P_{1/2,\pm 1/2,\pi/2}=P_{1/2,\pm 1/2}=1/2$ shown schematically in Figure~\ref{fig:SGE_half} is 
interpreted by comparison with photon polarization. The sum of all probabilities
for a state ket $|+x\rangle$ (or $|-x\rangle$) to collapse as the basis kets 
$|+z\rangle$ and $|-z\rangle$ adds to one. 
For a particular $m$, the total probability follows
\begin{eqnarray}
  P_{m}^{(j)} \left( \beta \right) = 
    \sum_{m^\prime } P_{m^\prime m}^{(j)} \left( \beta \right) = 1.
\end{eqnarray}
For $j=1$, the Wigner probabilities are schematically shown
in Figure~\ref{fig:spin_one}.


For integer values of $j$ only
$(j+1)^2$ terms are unique, and when $j$ takes half-integer values, only
$(j+1/2)(j+3/2)$ terms are unique due to the following symmetry relations
\begin{eqnarray}
P_{m^\prime m}^{(j)} \left( \beta \right)&=&
  P_{m m^\prime}^{(j)} \left( \beta \right) 
= P_{-m^\prime -m}^{(j)} \left( \beta \right)
= P_{-m -m^\prime}^{(j)} \left( \beta \right)    \\
P_{m^\prime -m}^{(j)} \left( \beta \right)&=&
  P_{-m m^\prime}^{(j)} \left( \beta \right).
\end{eqnarray}

Hence, the unique probabilities can be denoted compactly as equations rather than schematic diagrams as follows:
\begin{eqnarray}
j=1/2:&& 
P_{-1/2,-1/2}^{(1/2)}(\beta)=\left(\cos \beta/2 \right)^2,\, 
P_{-1/2,1/2}^{(1/2)}(\beta)=\left(\sin \beta / 2 \right)^2\\
j=1:&& 
P_{-1,-1}^{(1)}(\beta)=(\cos \beta/2)^4, \,
P_{-1,0}^{(1)}(\beta)=(\sin \beta)^2/2, \,
P_{-1,1}^{(1)}(\beta)=(\sin \beta/2)^4, \nonumber \\
&&
P_{0,0}^{(1)}(\beta)=(\cos \beta)^2  \\
j=3/2:&& 
P_{-3/2,-3/2}^{(3/2)}(\beta)=(\cos\beta/2)^6, \,
P_{-3/2,-1/2}^{(3/2)}(\beta)=3(\sin\beta/2)^2 (\cos \beta/2)^4, \nonumber \\
&&
P_{-3/2,1/2}^{(3/2)}(\beta)=3(\sin \beta/2)^4(\cos\beta/2)^2, \,
P_{-3/2,3/2}^{(3/2)}(\beta)=(\sin \beta/2)^6\, \nonumber \\
&&
P_{-1/2,-1/2}^{(3/2)}(\beta)=(2 (\sin\beta/2)^2 \cos\beta/2 - (\cos \beta/2)^3)^2,\, \nonumber \\
&&
P_{-1/2,1/2}^{(3/2)}(\beta)=((\sin \beta/2)^3 - 2\sin \beta/2 (\cos \beta/2)^2)^2  \\
j=2:&& 
P_{-2,-2}^{(2)}(\beta)=(\cos\beta/2)^8, \,
P_{-2,-1}^{(2)}(\beta)=4(\sin \beta/2)^2 (\cos \beta/2)^6, \, \nonumber \\
&&
P_{-2,0}^{(2)}(\beta)=6(\sin \beta/2)^4 (\cos \beta/2)^4, \,
P_{-2,1}^{(2)}(\beta)=4(\sin \beta/2)^6 (\cos \beta/2)^2, \, \nonumber \\
&&
P_{-2,2}^{(2)}(\beta)=(\sin \beta/2)^8, \,
P_{-1,-1}^{(2)}(\beta)=(3(\sin \beta/2)^2 (\cos \beta/2)^2 - (\cos \beta/2)^4)^2, \,\nonumber \\
&&
P_{-1,0}^{(2)}(\beta)=6((\sin\beta/2)^3 (\cos\beta/2) - (\cos\beta/2)^3(\sin\beta/2))^2,\, \nonumber \\
&&
P_{-1,1}^{(2)}(\beta)=((\sin \beta/2)^4 - 3(\sin \beta/2)^2 (\cos \beta/2)^2)^2, \,\nonumber \\
&&
P_{0,0}^{(2)}(\beta)=((\sin \beta/2)^4 - 4(\sin \beta/2)^2 (\cos \beta/2)^2 + (\cos \beta/2)^4)^2.
\end{eqnarray}

\begin{figure}[h]
\begin{center}
\includegraphics[width=1.0\linewidth]{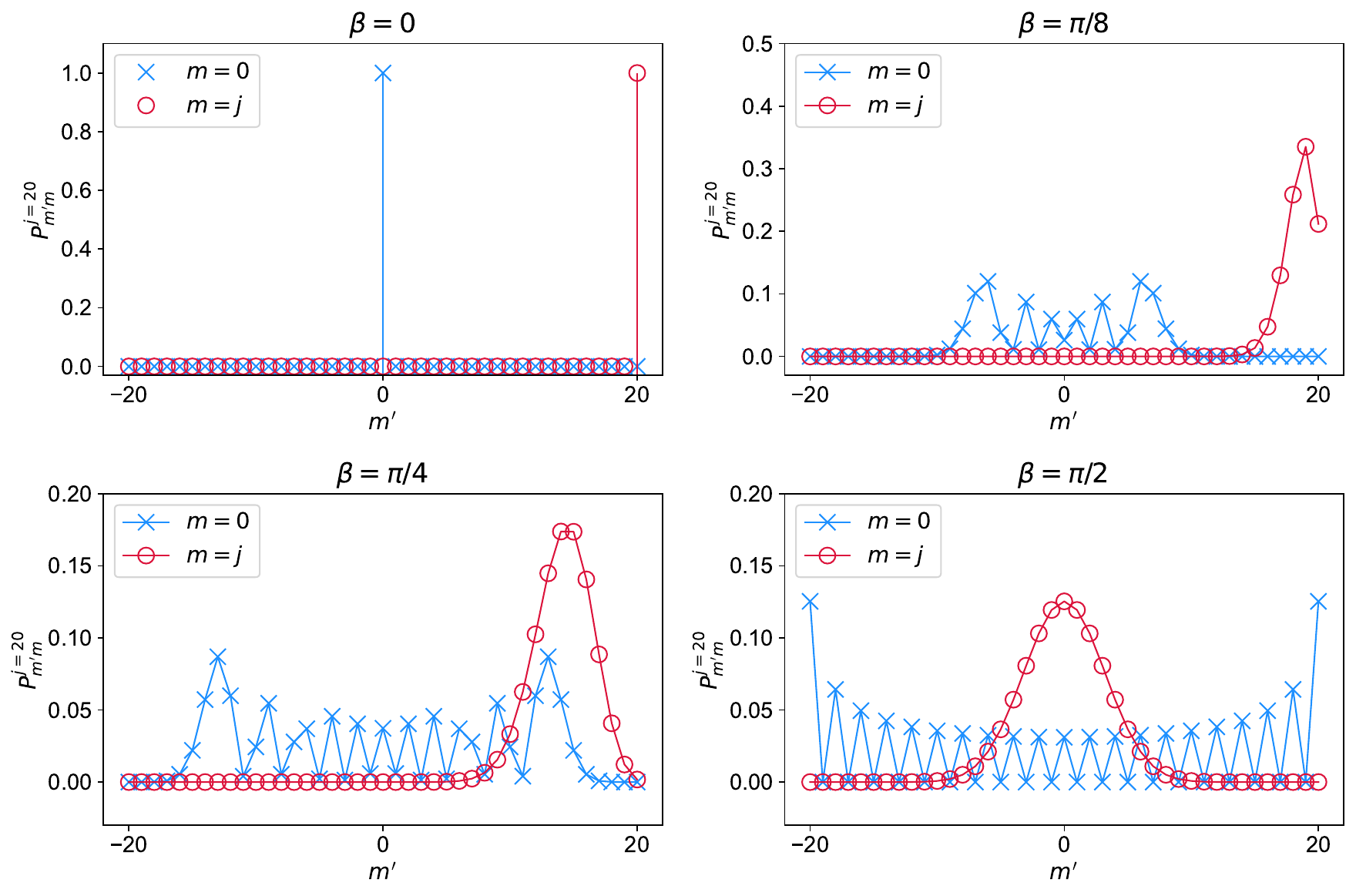}
\caption{
\footnotesize {Wigner probabilities for $j=20$.}
}
\label{fig:j_20}
\end{center}
\end{figure}

For large values of $j$, the probabilities can be plotted for various 
$\beta$ as shown in Figure~\ref{fig:j_20} 
dor the extreme cases of $m=0$ and $m=j$ for $j=20$. For these particular choices, 
Eq.~(\ref{eq:WignerFunction}) reduces to 
\begin{eqnarray}
    d_{m^\prime m=0}^{(j)} \left( \beta \right) & = &
    \sum_{k=\max(0,m-m^\prime)}^{\min(j+m,j-m^\prime)} (-1)^{k+m^\prime} 
    \frac
    {j!
    \sqrt{  \left(j + m^\prime \right)! \left(j - m^\prime \right)! }
    }
    {\left( j  - k \right)! k! \left( j - k - m^\prime  \right)! \left( k  + m^\prime  \right)! }  \nonumber \\
    & & \times
    \left( \cos \frac{\beta}{2}  \right)^{2j-2k-m^\prime}
    \left( \sin \frac{\beta}{2}  \right)^{2k+m^\prime}, 
    \label{eq:WignerFunction2_0}
\end{eqnarray}
and
\begin{eqnarray}
    d_{m^\prime m=j}^{(j)} \left( \beta \right) & = &
    \sum_{k=\max(0,m-m^\prime)}^{\min(j+m,j-m^\prime)} (-1)^{k-j+m^\prime} 
    \frac
    {
    \sqrt{ 2j!  \left(j + m^\prime \right)! \left(j - m^\prime \right)! }
    }
    {\left( 2 j  - k \right)! k! \left( j - k - m^\prime  \right)! \left( k - j + m^\prime  \right)! }  \nonumber \\
    & & \times
    \left( \cos \frac{\beta}{2}  \right)^{3j-2k-m^\prime}
    \left( \sin \frac{\beta}{2}  \right)^{2k-j+m^\prime}.
    \label{eq:WignerFunction2}
\end{eqnarray}

For $\beta=0$, the probabilities for a state kets collapse at
$m^\prime=m$ which is interpreted as the orthonormality
relations, $\langle j, m^\prime| j, m \rangle = \delta_{m,m^\prime}$.
When $m=0$, the probabilities are symmetrically distributed over positive and negative values of $m^\prime$ irrespective of the value of $\beta$. 
In general, when $\beta \ne 0$, the state kets $| j, m; \beta \rangle$ are non-trivial linear combinations of
$| j, m^\prime \rangle $ with Wigner probabilities spread across $m^\prime$
\begin{eqnarray}
 | j, m; \beta \rangle = 
    \sum_{m^\prime} | j, m^\prime  \rangle d_{m^\prime m}^{(j)} \left( \beta \right).
\end{eqnarray}
 In the rest of this note, we will focus on the classical limit of Wigner probabilities for $\beta=\pi/2$. 

\begin{figure}[!hbpt]
\begin{center}
\includegraphics[width=0.5\linewidth]{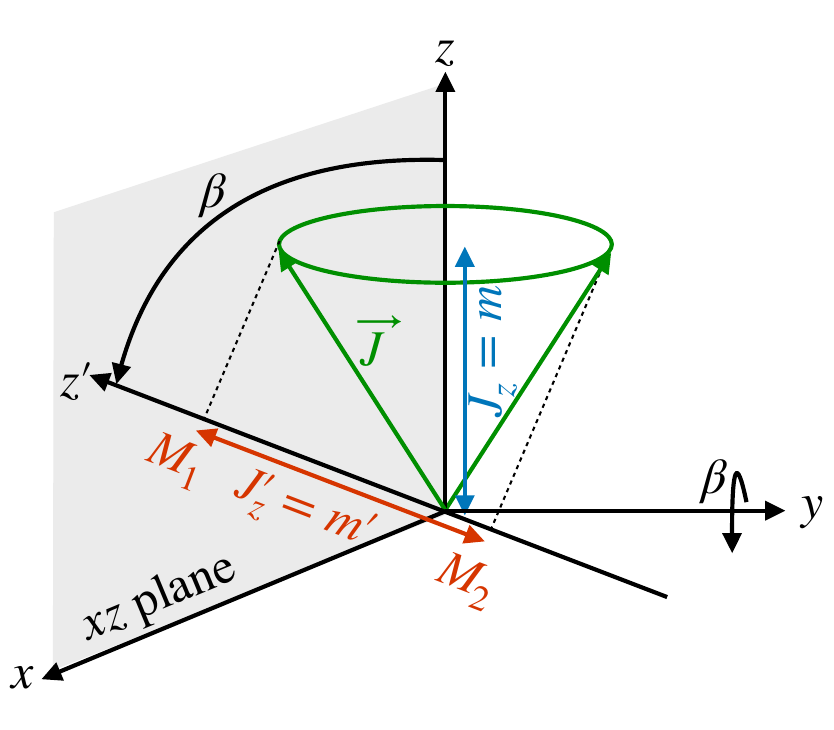}
\caption{
\footnotesize {Vector model interpretation of Wigner probabilities.}
}
\label{eq:ClassicalVector}
\end{center}
\end{figure}

For a classical angular momentum vector of length, $\vec{J}_{\rm cl.}$, the components
$J_x$, $J_y$, and $J_z$ can be simultaneously determined. 
For a quantum mechanical angular momentum vector $\vec{J}$ of length $\sqrt{ j(j+1) }$ (in units of $\hbar$) with the component along $z$-axis measured to be $m$, $J_x$ and $J_y$ components of $\vec{J}$ are indeterminate. To account for this indeterminacy, the vector model assumes $\vec{J}$ to precess around the axis of the external magnetic field (taken as the $z$-axis).  
The projection of the precessing vector along another 
axis, $z^\prime$,  will span a range of values  
$J_{z^\prime}=m^\prime \in \left[ M_1, M_2 \right]$ 
as shown in Figure~\ref{eq:ClassicalVector}. The corresponding
probability is interpreted as the classical limit of 
$P^{(j)}_{m^\prime m}(\beta)$\cite{BrussaardTolhoek,BrinkAngMom,Zare}.

\section{Local Averaging as a Consequence of Uncertainty in the Measurement of Position}
As the position basis kets form a complete basis, any state ket
can be written as a continuous superposition
of all the position basis kets
\begin{eqnarray}
    | \alpha \rangle = \int_{-\infty}^{+\infty}dx^{\prime\prime} | x^{\prime\prime} \rangle \langle x^{\prime\prime} | \alpha \rangle.
\end{eqnarray}
However, in the spirit of Heisenberg's thought experiment of $\gamma$-ray microscopy~\cite{Heisenberg}, 
a state ket upon position measurement cannot collapse to a basis set, $| x^{\prime\prime}\rangle$. The best one can observe is the formation of a 
localized superposition of position kets around 
$x^{\prime\prime}=x^{\prime}$~\cite{SakuraiQM} 
\begin{eqnarray}
     | \alpha \rangle & \xrightarrow[]{\,{\rm measurement}\,} &
    \int_{x^{\prime}-\Delta / 2}^{x^{\prime}+\Delta / 2}dx^{\prime\prime}\, | x^{\prime\prime} \rangle \langle x^{\prime\prime} | \alpha \rangle
    \label{eq:PositionUncertainty1}
\end{eqnarray}
 or considering normalization, as the local average 
\begin{eqnarray}
     | \alpha \rangle & \xrightarrow[]{\,{\rm measurement}\,} &
     \frac{1}{\Delta}
    \int_{x^{\prime}-\Delta / 2}^{x^{\prime}+\Delta / 2}dx^{\prime\prime}\, | x^{\prime\prime} \rangle \langle x^{\prime\prime} | \alpha \rangle.
    \label{eq:PositionUncertainty2}
\end{eqnarray}
The parameter $\Delta$ is the uncertainty in the measurement of position, 
which according to state-of-the-art high-resolution transmission electron microscopy is slightly below $0.5$ \AA\cite{ErniAtomicResolution}. Extending Eq.(\ref{eq:PositionUncertainty2})
to the measurement of the probability density at a given value of 
$x=x^\prime$ results in a local average of probability density 
around $x^\prime$ 
\begin{eqnarray}
|\langle x^\prime  | \alpha \rangle |^2 & \xrightarrow[]{\,{\rm measurement}\,} &
\frac{1}{\Delta}
    \int_{x^{\prime}-\Delta / 2}^{x^{\prime}+\Delta / 2}dx^{\prime\prime}\, 
    |\langle x^{\prime\prime} | \alpha \rangle |^2
     \label{eq:PositionUncertainty3}
\end{eqnarray}
Defining the wavefunction for the state ket as 
$\psi_\alpha(x^{\prime\prime})=\langle x^{\prime\prime} | \alpha \rangle$, we can 
write Eq.~(\ref{eq:PositionUncertainty3}) as
\begin{eqnarray}
    \langle P_{\rm QM}\rangle (x^\prime) & = & \frac{1}{\Delta} 
    \int_{x^{\prime}-\Delta/2}^{x^{\prime}+\Delta/2} dx^{\prime\prime} 
    | \psi_\alpha(x^{\prime\prime}) |^2.
    \label{eq:local_average2}
\end{eqnarray}

\subsection{Particle-in-a-box and Simple harmonic oscillator:}
\begin{figure}[h]
\begin{center}
\includegraphics[width=\linewidth]{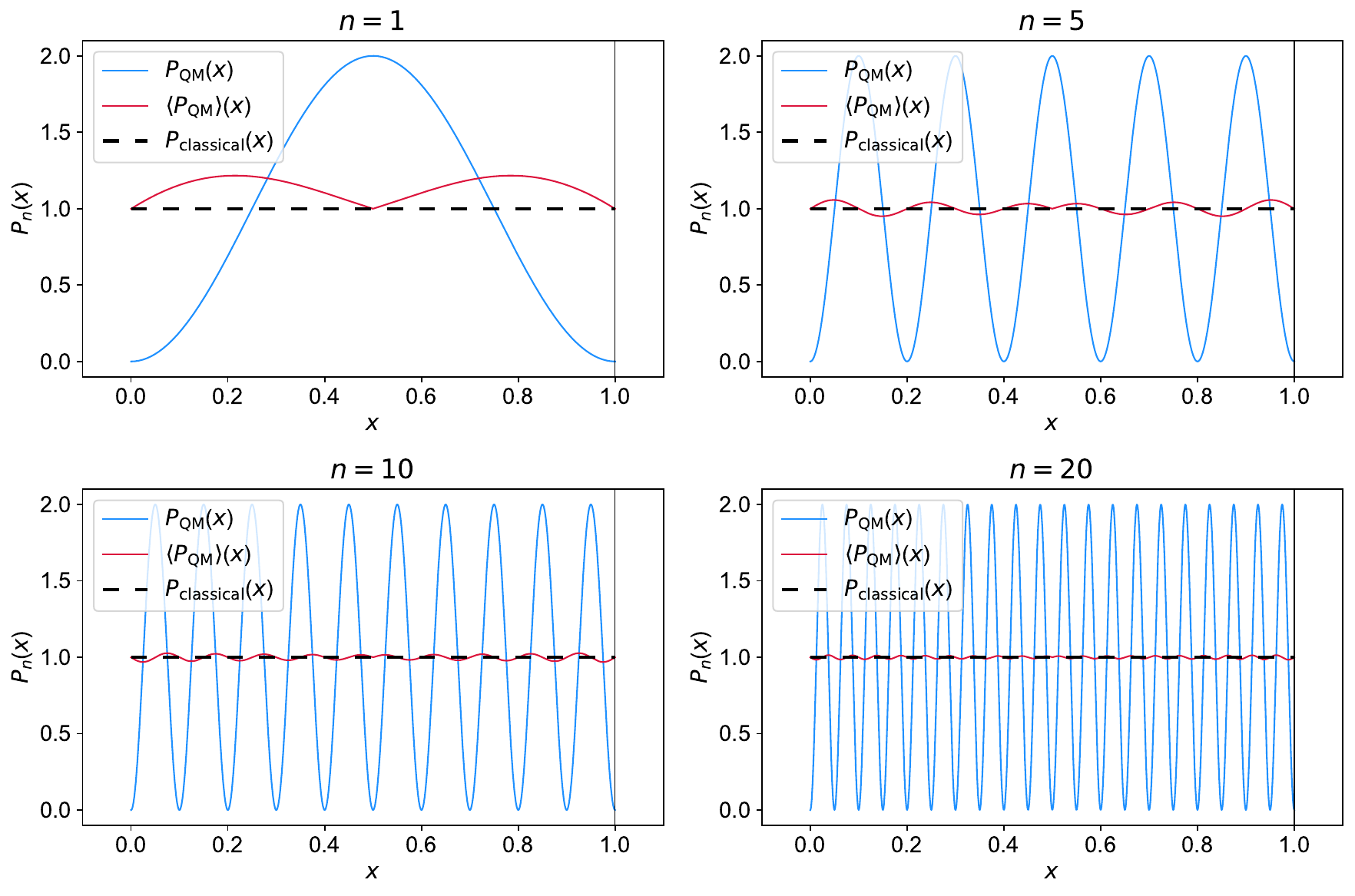}
\caption{
\footnotesize {
Probability densities for particle-in-a-box of length, $L=1$. 
Local averaging of quantum mechanical probability density was 
performed using Eq.~\ref{eq:local_average2} with $\Delta=1$.
The dashed curve denotes the classical probability density $1/L$.
}}
\label{fig:PIB}
\end{center}
\end{figure}

\begin{figure}[h]
\begin{center}
\includegraphics[width=\linewidth]{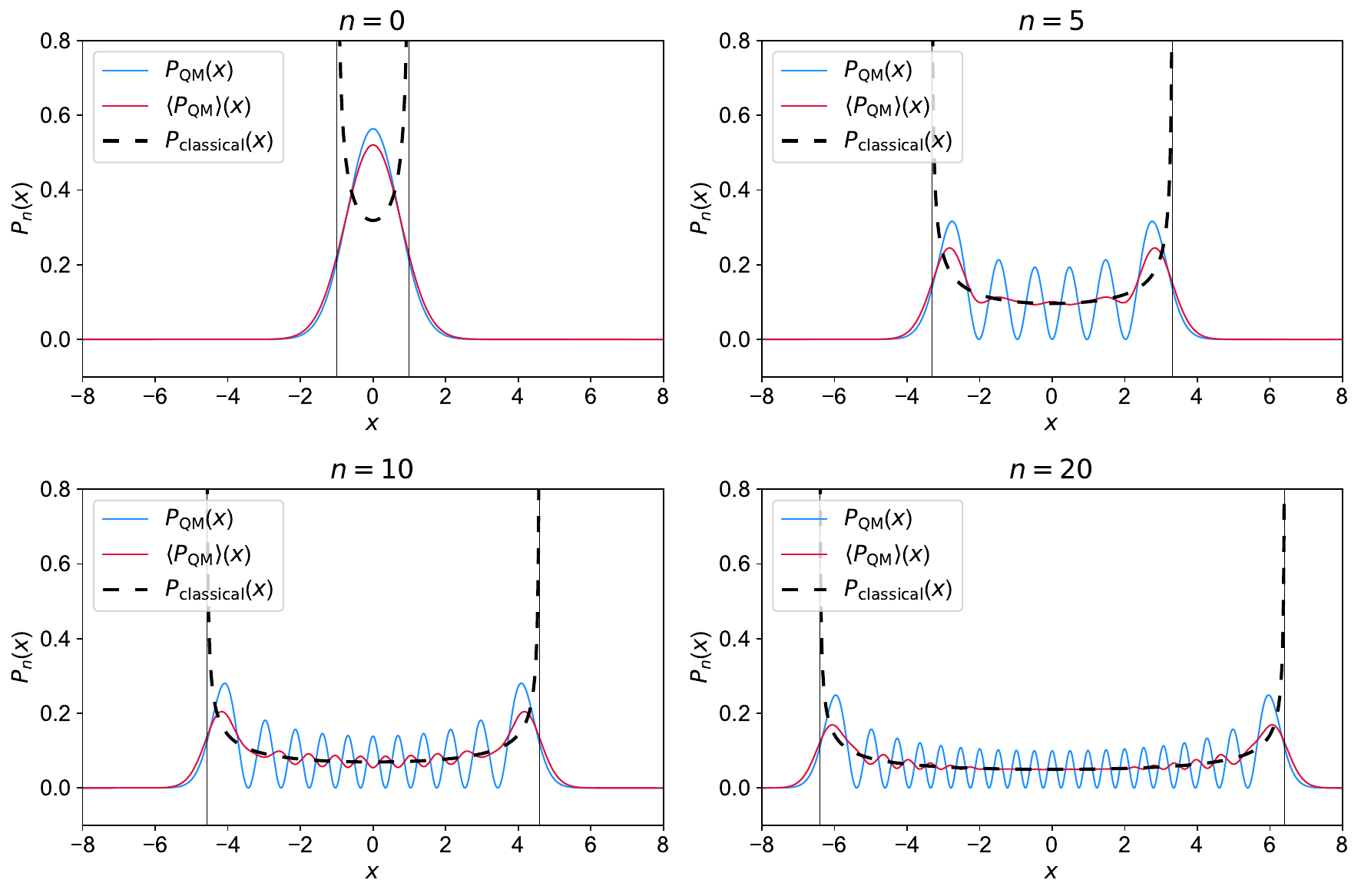}
\caption{
\footnotesize {
Probability densities for simple harmonic oscillator ($m=\omega=\hbar=1$). 
% CFP denotes the classically forbidden probability 
% for locating the particle beyond the classical turning points shown in vertical lines. 
Local averaging of quantum mechanical probability density was performed using $\Delta=1$.
The dashed curve denotes the classical probability density calculated using the energy, $E_n=\hbar \omega (n+1/2)$.
}
}
\label{fig:SHO}
\end{center}
\end{figure}

The limiting value of the quantum number for which the locally averaged 
probability density approaches the classical result
depends on $\Delta$, which is the minimum distance with 
which position kets can be resolved.  
For the quantum mechanical models, particle-in-a-box (PIB) and
the simple harmonic oscillator (SHO), locally averaged quantum 
 mechanical probabilities using 
$\Delta=1$ (in atomic units, which is 0.529 \AA) essentially 
removes fluctuations for $n\ge20$
as shown in Figure~\ref{fig:PIB} and Figure~\ref{fig:SHO}. 

The classical limit of PIB is a particle confined in 
$0< x <L$ traveling back and forth across the box
at a great speed,  for which the probability of locating a particle 
between $x$ and $x+dx$ is the uniform distribution, $P_{\rm classical}^{\rm PIB} (x) dx=dx/L$.
The classical probability of locating a particle bound by a harmonic potential
between $x$ and $x+dx$ is inversely
proportional to the velocity of the particle at that position.
\begin{eqnarray}
    P_{\rm classical}^{\rm SHO} (x) dx \propto \frac{1}{v(x)} dx
\end{eqnarray}
Using the fact that the sum of kinetic and potential energy is conserved, $v$
can be expressed as a function of the classical turning point, $x_t$
\begin{eqnarray}
  E = \frac{mv^2}{2} + \frac{m\omega^2x^2}{2} = \frac{m\omega^2 x_t^2}{2}; \quad
    P_{\rm classical}^{\rm SHO} (x) dx \propto \frac{1}{\omega \sqrt{x_t^2 - x^2}}  dx 
\end{eqnarray}
The proportionality constant can be determined by considering that the
total probability is 1.
\begin{eqnarray}
    P_{\rm classical}^{\rm SHO} (x) dx  = \frac{1}{ \pi \omega \sqrt{x_t^2 - x^2}}  dx 
\end{eqnarray}
If a smaller value of $\Delta$ 
is considered, then the classical limit
will be approached for a higher quantum number. In principle, if one considers the limit of $\Delta \rightarrow 0$, the classical limit
is reached only when the quantum number approaches infinity.

\subsection{Classical limit of Wigner probabilities:}
\begin{figure}[h]
\begin{center}
\includegraphics[width=1.0\linewidth]{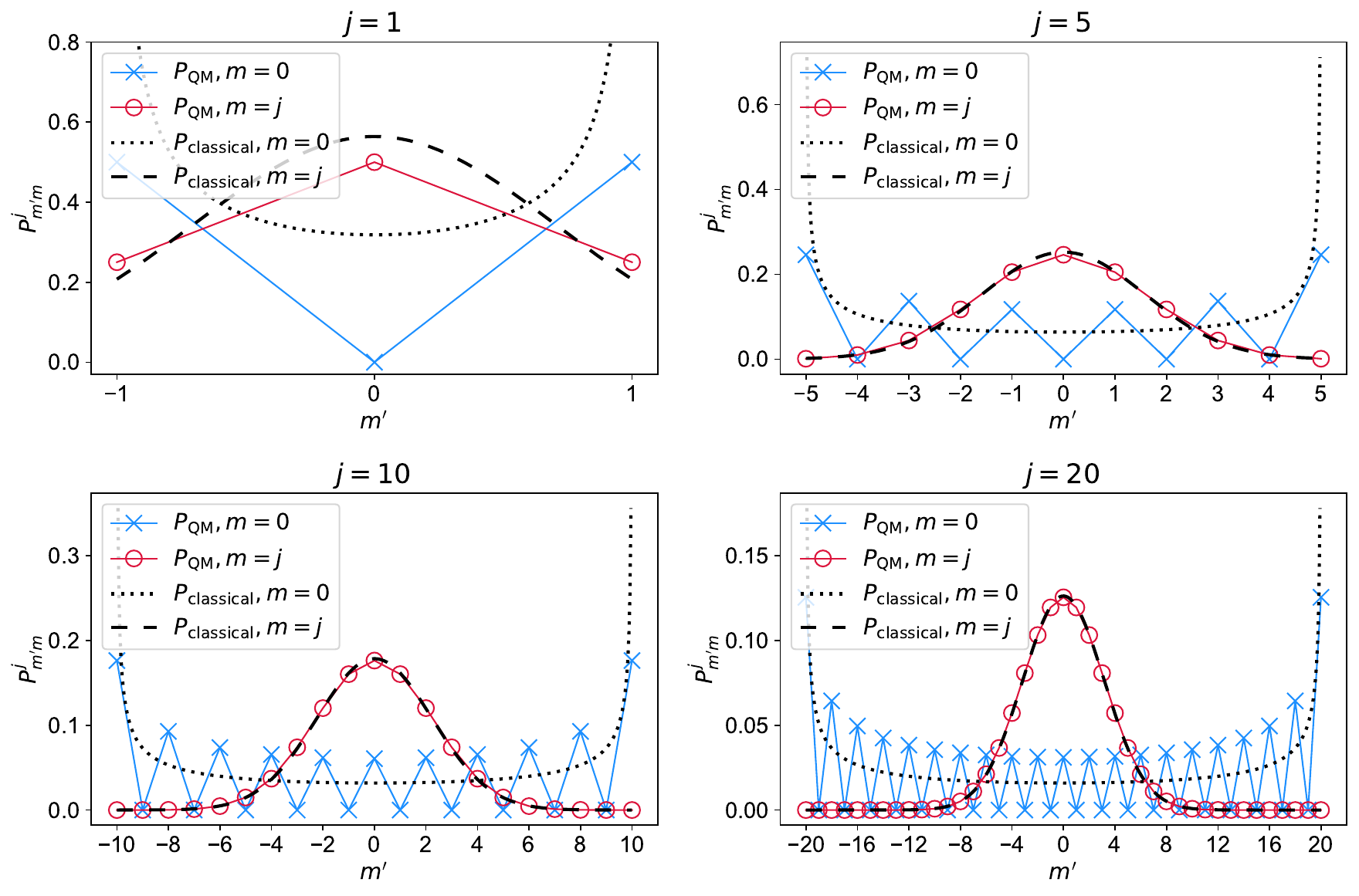}
\caption{
\footnotesize {Quantum and classical probabilities when 
$| j, m; \beta = \pi/2\rangle $ is 
$| j, 0; \pi/2 \rangle $ or $| j, j; \pi/2 \rangle $.}
}
\label{fig:beta_pi_by_2}
\end{center}
\end{figure}

For the special case of $\beta=\pi/2$, Wigner probabilities given in
Eq.~(\ref{eq:WignerFunction2_0}) and 
Eq.~(\ref{eq:WignerFunction2}) reduce to 
\begin{eqnarray}
    d_{m^\prime m=j}^{(j)} \left( \beta=\pi/2 \right) & = &
    \sum_{k=\max(0,j-m^\prime)}^{\min(2j,j-m^\prime)} 
    \frac
    {(-1)^{k-j+m^\prime}
    \sqrt{ 2j!  \left(j + m^\prime \right)! \left(j - m^\prime \right)! }
    }
    {2^j\left( 2 j  - k \right)! k! \left( j - k - m^\prime  \right)! \left( k - j + m^\prime  \right)! },
\end{eqnarray}
and 
\begin{eqnarray}
        d_{m^\prime m=0}^{(j)} \left( \beta=\pi/2 \right) & = &
    \sum_{k=\max(0,m^\prime)}^{\min(j,j-m^\prime)}
    \frac
    {(-1)^{k+m^\prime} j!
    \sqrt{  \left(j + m^\prime \right)! \left(j - m^\prime \right)! }
    }
    {2^j\left( j  - k \right)! k! \left( j - k - m^\prime  \right)! \left( k  + m^\prime  \right)! }  .
    \label{eq:WignerFunction3}
\end{eqnarray}
Probabilities determined using the above equations are plotted in Figure~\ref{fig:beta_pi_by_2} for selected values of $j$. The 
spread of $P_{\rm QM}(m^\prime,m;\beta=\pi/2)$ over a range of $m^\prime$ 
ensures that the uncertainty relation
 $\Delta J_x \Delta J_z \ge \hbar |\langle J_y \rangle |/2$ is obeyed. For example,
if $J_z$ is known to have the value $m$ with arbitrary precision, 
 $J_{z^{\prime}}=J_x$ can be measured only with a finite precision.
 The classical limit, according to the processing vector model, is the probability that 
$J_{z^\prime}$ will take a particular value $m^\prime$~\cite{BrinkAngMom} 
\begin{eqnarray}
P_{m^\prime m}^{(j), {\rm classical}} \left( \beta \right) = \frac{1}
    { \pi \sqrt{j^2 (1-\cos^2\beta) - \left( m\hat{z}-m^\prime \hat{z}^\prime \right)^2 } }.
    \label{eq:BS_1}
\end{eqnarray}

%m=j
\subsection{$P_{m^\prime, m=j}^{(j)}\left( \beta=\pi/2 \right)$:}
Eq.~(\ref{eq:BS_1}) with $\beta=\pi/2$ and $m=j$ corresponds to a singularity 
 at $m^\prime=0$. The corresponding quantum mechanical 
 result shown in Figure~\ref{fig:beta_pi_by_2} for $j=20$ 
 is described by the Gaussian distribution:
\begin{eqnarray}
    P_{\rm classical}(m^\prime,m=j;\beta=\pi/2) & = & \frac{1}{\sqrt{2 \pi \sigma^2}} \exp \left( 
    -\frac{(m^\prime-\mu)^2}{2 \sigma^2}\right), 
    \label{eq:Gaussian}
\end{eqnarray}
centered at $m^\prime=\mu=0$ with standard deviation $\sigma=\sqrt{j/2}$. 
The Gaussian function approaches the Dirac delta function 
at the limit $j\rightarrow 0$, which is not the classical limit. However, 
comparing probability densities for different values of $j$, the variable 
$m^\prime$ has to be replaced by $M =m^\prime/j$ resulting in
\begin{eqnarray}
    P_{\rm classical}(M,m=j;\beta=\pi/2) & = & \sqrt{\frac{j}{\pi}} \exp \left( -jM^2\right), 
    \label{eq:Gaussian2}
\end{eqnarray}
which for the classical limit $j\rightarrow \infty$ represents the Dirac delta function. 
The shape of $P_{\rm classical}(M,m=j;\beta=\pi/2)$ can be interpreted by
comparison with the probability density of SHO with $n=0$, where
the uncertainties in position and momentum follow $\Delta x \Delta p \ge \hbar/2$. Hence, the state with the particle localized at $x=0$ will imply infinite $\Delta p$.

%m=0
\subsection{$P_{m^\prime, m=0}^{(j)}\left( \beta=\pi/2 \right)$:}
\begin{figure}[!hbpt]
\begin{center}
\includegraphics[width=0.5\linewidth]{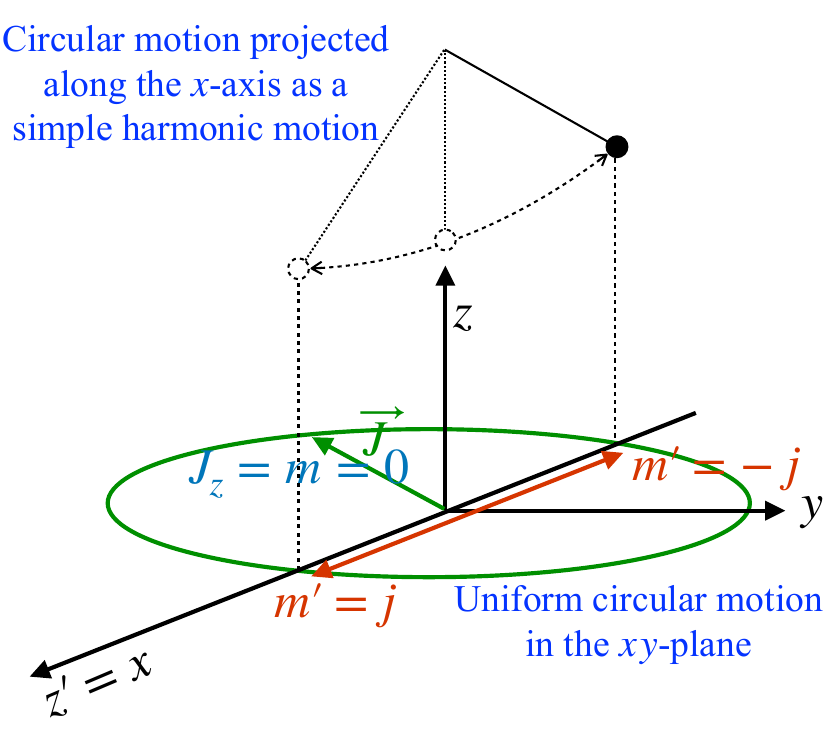}
\caption{
\footnotesize {Vector model interpretation of the precession of $\vec{J}$
in the $xy$-plane with $J_z=m=0$.}
}
\label{eq:ClassicalVectorxy}
\end{center}
\end{figure}

The more interesting case is that of $m=0$, for which
$P_{\rm QM}(m^\prime)$ shows fluctuations. 
For $\beta=\pi/2$ and $m=0$, Eq.~\ref{eq:BS_1} reduces to
\begin{eqnarray}
    P_{\rm classical}(m^\prime,m=0;\beta=\pi/2) = \frac{1}
    { \pi \sqrt{j^2 - {m^\prime}^2 } }.
    \label{eq:BS_2}
\end{eqnarray}
Figure~\ref{fig:beta_pi_by_2} suggests the classical limit (large $j$) 
to agree with Eq.~(\ref{eq:BS_2}). The reason why the classical limit for the 
spin-problem resembles that of the classical limit for SHO, is  
the uniform circular motion traced by the precessing spin vector with the angular velocity of precession ($\omega_p$) in the $xy$-plane projected along
either of the axes is a simple harmonic motion with angular
velocity, $\omega=\omega_p$, as illustrated in Figure~\ref{eq:ClassicalVectorxy}. 
 Another resemblance with a well-known result is that the nodes (zero-probabilities) for $P_{\rm QM}(m^\prime,m=0)$ are equally spaced as in the case of the
particle-in-a-box problem, while the nodes of the harmonic oscillator wavefunctions are distributed densely near the origin and sparsely 
near the turning points. However, local averaging affects with a suitable resolution
the distribution of nodes equally resulting in the same classical limit.

In the classic SGE, the magnitude of the component of the spin angular momentum of silver (Ag)
atom is inferred from the spatial separation of Ag atoms deposited on a glass plate. Newtonian mechanics is applied to the
atom's trajectory through the inhomogeneous magnetic field. 
When applying the standard values used for this problem\cite{Halliday}---$g=2$ (g-factor for electron),
$e=1.6\times10^{-19}$ C (charge of an electron),
$M_{\rm Ag}=1.8\times10^{-25}$ kg (mass of silver atom),
$M_{\rm e}=9.1\times10^{-31}$ kg (mass of electron),
$\left| \frac{\partial B_z}{\partial z}\right|$ = 1400 T/m (magnetic field gradient),
$L=3.5$ cm (length of the magnet), and
$v=750$ m/s (mean speed of Ag atoms from the furnace), along with 
$s_z=\hbar/2$ (the magnitude of the spin-angular momentum along the magnetic field)---the separation of adjacent peaks ($\Delta w$) amounts to $2w=0.16$ mm, where $w$ is
\begin{eqnarray}
    w & = & -\frac{ge}{4 M_{\rm Ag}m_{\rm e}}
    \left| \frac{\partial B_z}{\partial z}\right| 
    \left( \frac{L}{v}\right)^2 s_z.
\end{eqnarray}

$\Delta w$ in the SGE is independent of the 
spin quantum number. Hence, for microscopic systems such as an
electron or an atom, Wigner probabilities at 
the limit of large quantum numbers cannot be thought of as approaching the classical limit.
However, if the silver atoms are replaced by particles heavier
by a factor of $10^6$ (or particles 100 times heavier than a virus), 
$\Delta w$ decreases to $\mathcal{O}\left( {\rm \AA} \right)$. In such cases, 
the oscillatory behavior in the Wigner probabilities for $j=\mathcal{O}(10)$ can be locally averaged over 
$m^\prime$ values at a resolution of $\Delta m^\prime = 1$. 
The resulting locally averaged probabilities, $\langle P_{\rm QM}(m^\prime,m=0) \rangle $ and 
$\langle P_{\rm QM}(m^\prime,m=j) \rangle $,
will resemble  
$P_{\rm QM}^{\rm SHO} (n=0)$ and 
$\langle P_{\rm QM}^{\rm SHO}\rangle(n=\mathcal{O}(10))$, respectively.

\section{Conclusion\label{Conclusion}} 
The probabilistic outcomes in the Stern--Gerlach-type thought experiments for the simplest case of $j=1/2$ suggest that a system prepared in the $| \pm x \rangle $ state ket entering an inhomogeneous magnetic field along the $z$-axis can be found in the $| \pm z \rangle $ basis kets with
equal probability. 
The Wigner probabilities express this observation as follows:
$
P_{-1/2,-1/2}^{(1/2)}(\beta=\pi/2)=
P_{+1/2,-1/2}^{(1/2)}(\beta=\pi/2)=
P_{-1/2,+1/2}^{(1/2)}(\beta=\pi/2)=
P_{+1/2,+1/2}^{(1/2)}(\beta=\pi/2)=
1/2$. 
For $j=1$, the probability for the state ket 
$|j=1,m=\pm1;\beta=\pi/2\rangle$ to be found in the basis kets $|j=1,m^\prime \rangle$ is small for the extreme values $m^\prime=\pm 1$ and large for the central value $m^\prime= 0$. 
Conversely, the probability of the state ket 
$|j=1,m=0;\beta=\pi/2\rangle$ to be found in the basis kets $|j=1,m^\prime \rangle$ is large for the 
extreme values $m^\prime=\pm 1$ and small at $m^\prime= 0$. With increasing $j$, these
probabilities resemble those of the 
well-known SHO model. 
For $m^\prime= 0$, the position
of the nodes, {\it i.e.,} the values of $m^\prime$ for which the probabilities vanish become equally spaced, akin to that 
of another well-known model problem, the particle-in-a-box.  
If the initial states form a thermal ensemble, as in the case of 
the classic SGE, the resulting outcomes 
correspond to the
total probabilities, $P_{m}^{(j)} \left( \beta \right)$, 
summed over all values of $m^\prime$, yielding a uniform distribution of height 
$1/(2j+1)$. 
For spatial resolutions around $\approx{\rm \AA}$, beams corresponding to 
adjacent $m$-values become indistinguishable for 
$j$ as low as $\mathcal{O}(10)$ in systems with particle masses 
$\approx10^{-16}$ g for
magnetic field gradients of approximately $\approx10^3$ T/m.  

\section*{Acknowledgments} 
The author acknowledges the support of the 
Department of Atomic Energy, Government
of India, under Project Identification No.~RTI~4007.

\end {document}